\documentclass[twocolumn,aps,pre,floatfix]{revtex4}
\usepackage{graphicx,color}
\usepackage{amsmath,amssymb}

\newcommand{\hlat}{%
\begin{figure}[htbp]
   \centering
   \includegraphics[width=0.45\textwidth]{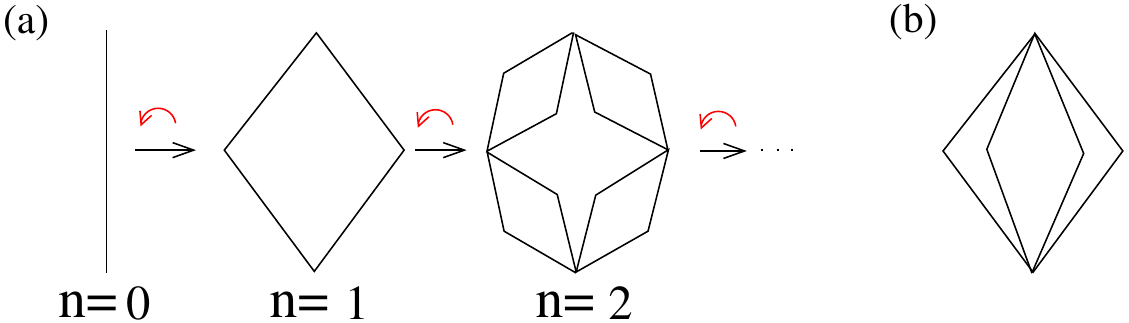}

   \caption{(Color online) (a) The recursive construction of the hierarchical 
    lattice with $b=2$ for $n=0,1,2,...$ generations.  The right arrows represent 
    the direction of iteration towards larger lattices.  The left arrows represent 
    the direction of decimation used in the RG.  (b) A motif of $2b$ bonds, where 
    $b=4$.   
}
   \label{fig:fig1}
 \end{figure}
}
\newcommand{\hlatint}{%
\begin{figure}[htbp]
   \centering
  \includegraphics[width=0.35\textwidth]{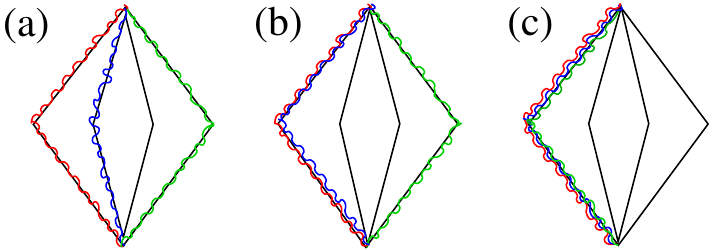}

   \caption{(Color online) Examples of three-chain configurations on a diamond 
    motif for $b=4$.  (a) The polymers do not share any single bond.  The number of 
    such configurations is $b(b-1)(b-2)$.  (b)Two polymers share a bond.  The 
    energy here is $-2\epsilon$ and the number of such configurations is $b(b-1)$.  
    (c) Three polymers share the same bond.  The energy is 
    $2(-3\epsilon-\epsilon_{123})$.  The number of such configurations is $b$.  }

   \label{fig:fig2}
 \end{figure}
}

\newcommand{\bfderridup}{%
\begin{figure}[htbp]
   \centering
   \includegraphics[width=0.45\textwidth]{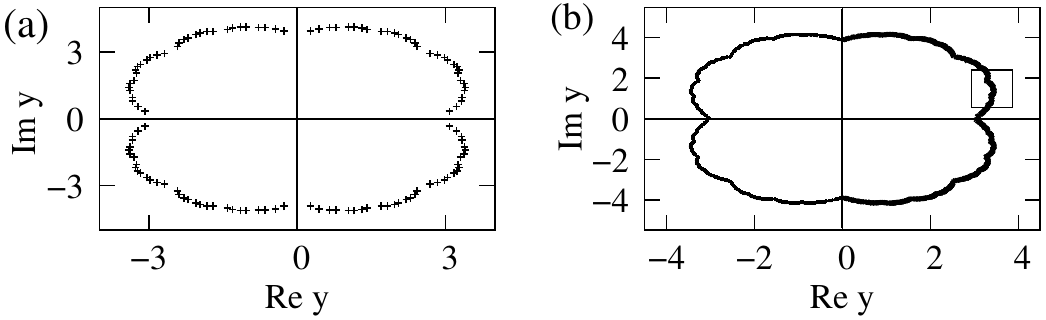}

   \caption{Plot of zeros of $Z_n(y)$ in the complex $y$-plane for $b=4$ from 
    (a) the exact recursion relation for $n=6$, and (b) the RG relation.  The 
    closest point 
    to the Re$(y)$ axis approaches $y_c=3$, the two-chain melting point in the 
    limit $n\to\infty$, the unstable fixed point of Eq.~(\ref{eq:ry}).  The 
    selected region shown by a box is zoomed in Fig.~\ref{fig:fig4}(a).
}

   \label{fig:fig3}
 \end{figure}
}

\newcommand{\figbfourfrac}{%
\begin{figure}[htbp]
   \centering
   \includegraphics[width=0.45\textwidth]{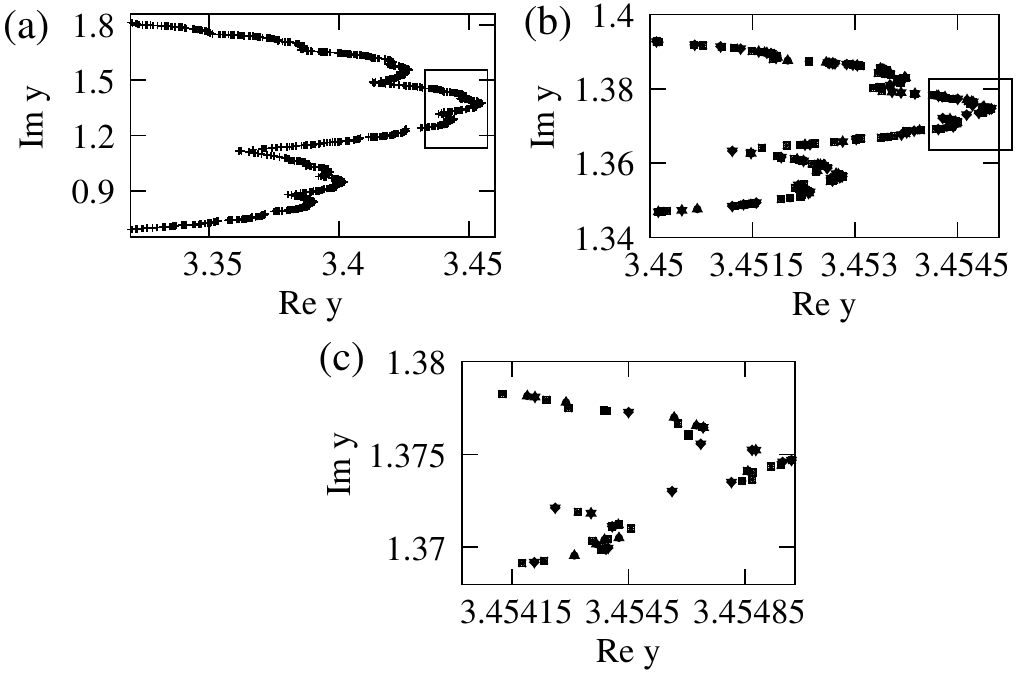}

   \caption{Zeros of $Z_n(y)$: The inner rectangular box is zoomed successively.  
   A self-similar structure becomes apparent.  Note that the zeros are known with 
   high accuracy.}

   \label{fig:fig4}
 \end{figure}
}

\newcommand{\bfdupflow}{%
\begin{figure}[htbp]
   \centering
   \includegraphics[width=0.45\textwidth]{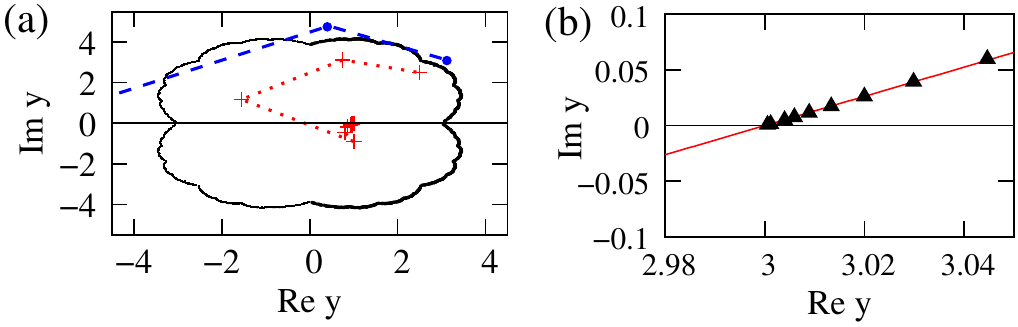}

   \caption{(Color online) (a) Plot of zeros of $Z_n(y)$ in the complex $y$ plane.  
    Two types of RG flow are shown.  The dotted red curve starts from a point of 
    the inner region and flows to $y=1$.  The dashed blue curve starts from a point 
    of the outer region and flows to $\infty$.   
    (b) The triangles are the zeros and approach the limit point $y_c=3$ at large 
    $n$.  The solid red line, given by Eq.~(\ref{eq:line}), makes an angle 
    $\phi$ with the real axis with  $\nu$ of Eq.~(\ref{eq:nuy}) and $c=y_c$. 
}

   \label{fig:fig5}
 \end{figure}
}

\newcommand{\figbfour}{%
\begin{figure}[htbp]
   \centering
   \includegraphics[width=0.45\textwidth]{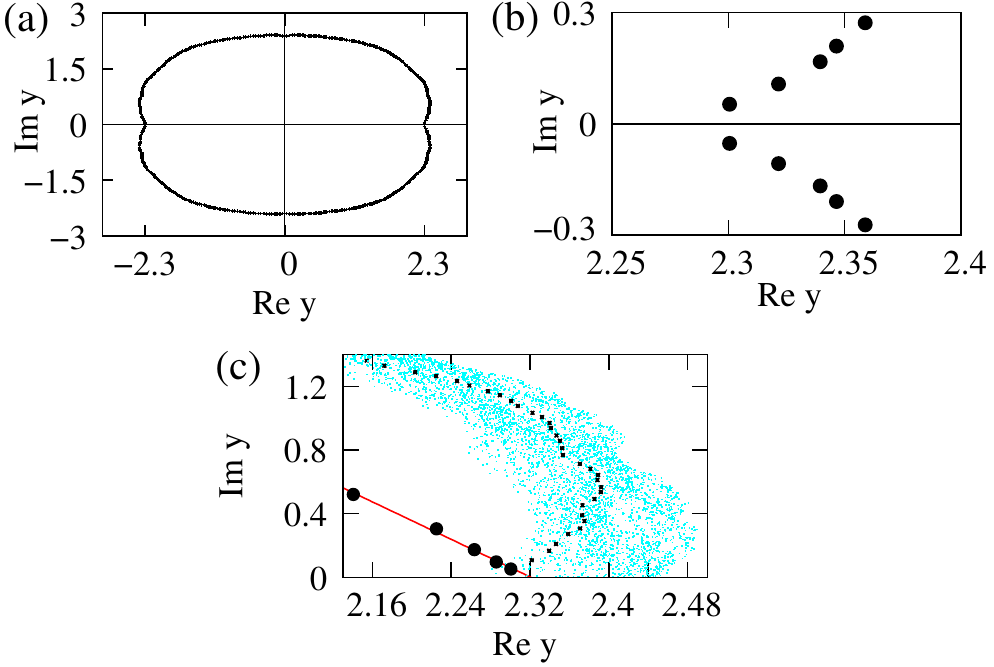}

   \caption{(Color online) Plot of zeros in the complex $y$-plane for $b=4$.
     (a) Zeros of $Q_n(y,1)$, when $n=6$, (b) a finer scale of (a) near the 
     real axis, and (c) combined plot of zeros.  The bigger black circles are the 
     zeros closest to the real axis ({\it{i. e.},} with smallest imaginary part) 
     obtained from $Q_n(y,1)$ for $n=2,...,6$ and the solid (red) straight line 
     is a fit to these.  The ``Milky Way"-like region shows the distribution of 
     zeros from Eq.~({\ref{eq:w1}}) on which we superpose the positive quadrant of
     (a) shown by the small black dots.  }

   \label{fig:fig6}
 \end{figure}
}

\newcommand{\bftripflow}{%
\begin{figure}[htbp]
   \centering
   \includegraphics[width=0.45\textwidth]{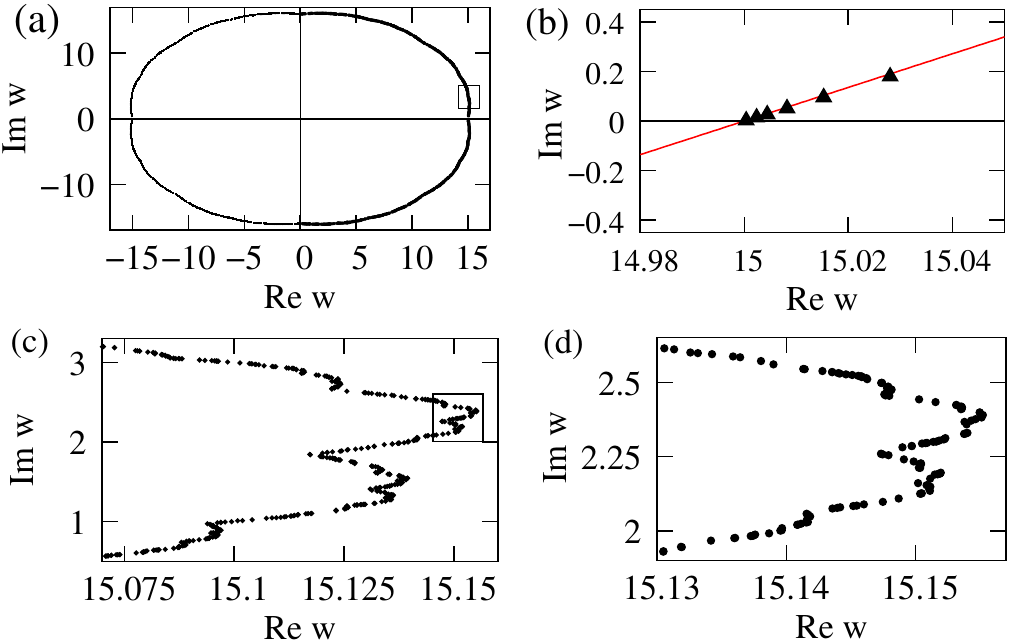}

   \caption{(Color online) (a) Plot of the zeros of $Q_n(1,w)$ in the complex 
    $w$-plane for $b=4$.  The closest point to the real axis approaches $w_c=15$ for 
    large $n$.  There is self-similarity in the distribution of zeros.  (b) The 
    triangles are the zeros.  The solid red line given by Eq.~(\ref{eq:line}) 
    passes through them with $\nu$ of Eq.~(\ref{eq:nuw}) and $c=w_c$.  
    (c), (d) The inner rectangular box [from (a)] is zoomed successively.             
    }

   \label{fig:fig7}
 \end{figure}
}

\newcommand{\figbnine}{%
\begin{figure}[htbp]
   \centering
   \includegraphics[width=0.45\textwidth]{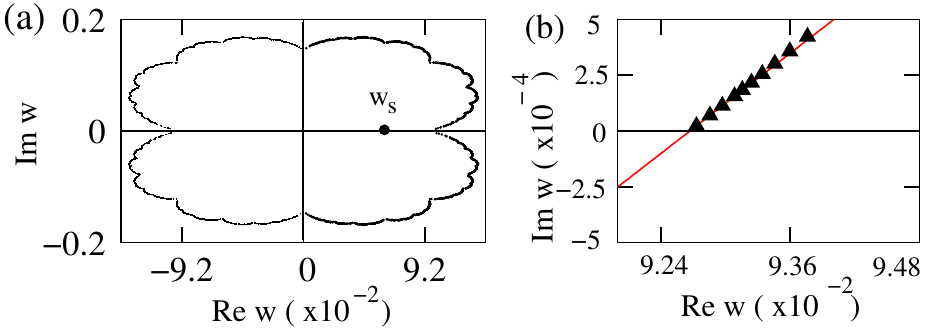}

   \caption{(Color online) (a) Zeros of $Q_n(y_c,w)$ in the complex $w$ plane 
    for $b=9$.  The stable fixed point $w=w_s$ is shown by a black circle.    
    (b) The solid red line is given by Eq.~(\ref{eq:line}) and passes through 
    the zeros shown by the triangles, with $\nu$ from Eq.~(\ref{eq:nuyc})
    and $c=w_E$.  
    }

   \label{fig:fig8}
 \end{figure}
}

\newcommand{\phasedia}{%
\begin{figure}[htbp]
   \centering
   \includegraphics[width=0.45\textwidth]{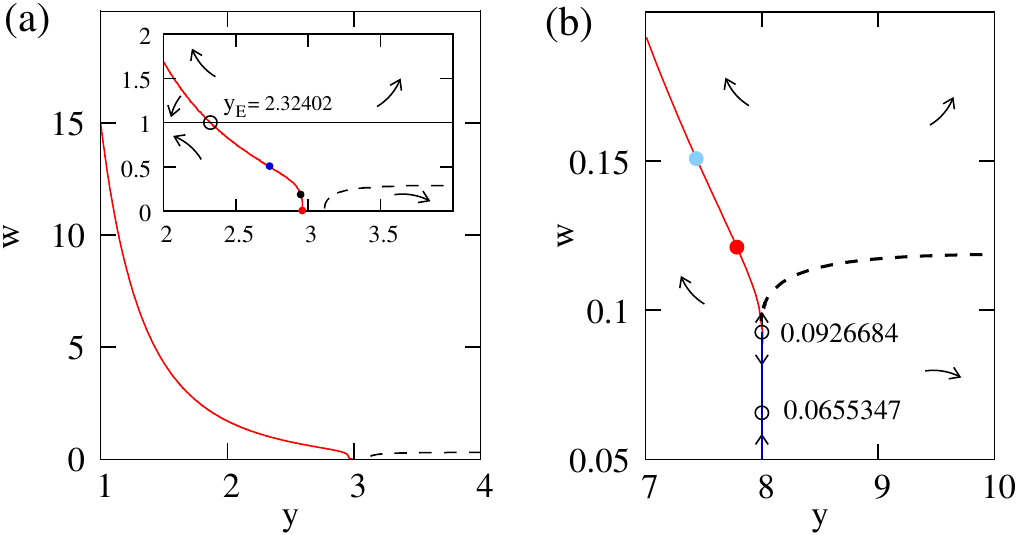}

   \caption{(Color online) RG phase diagram in the $y$-$w$ plane.  The arrows are to 
    show the flow of the renormalized parameter schematically.  (a) For $b=4$.  The 
    solid red curve and the dashed curve represent the separatrices, where $w$ 
    flows to two  different fixed points on either side of the separatrix.  The 
    zoomed area near $y_c=3$ is shown in the inset.  For $w=1$, $y_E=2.32402$ is 
    the Efimov DNA transition point.  The filled circles are the Efimov transition 
    points for $w=0.5$, and $w=0.2$, $w=0$, respectively, obtained from 
    Fig.~\ref{fig:fig12}(a).  (b) For $b=9$.  Along $y_c=8$, there are two real fixed 
    points given by Eq.~(\ref{eq:ws}).  The solid (red) and dashed lines are the 
    separatrices.  The filled circles are the Efimov transition points for 
    $w=0.15$ and $w=0.12$, respectively, obtained from Fig.~\ref{fig:fig12}(b).  
}

   \label{fig:fig9}
 \end{figure}
}

\newcommand{\tripchainnine}{%
\begin{figure}[htbp]
   \centering

   \includegraphics[width=0.45\textwidth]{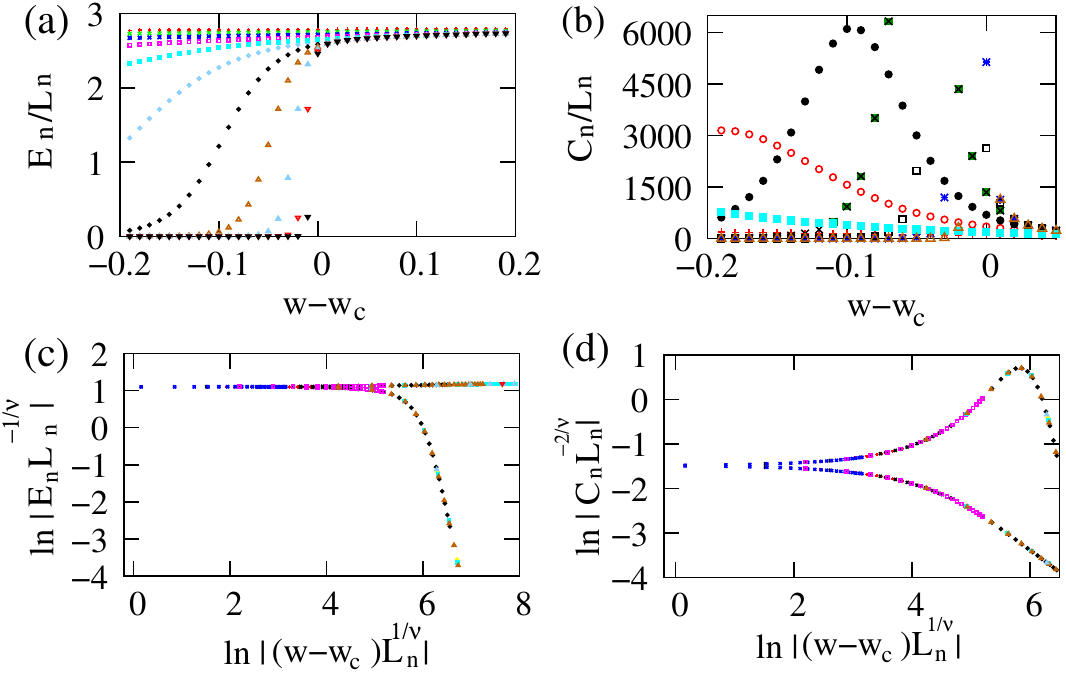}

   \caption{(Color online) For $b=4$.  (a) The three-chain average energy per 
    monomer versus the corresponding Boltzmann factor for chain length up to 
    $2^{26}$ when $y=1$.  The average energy shows a continuous transition at 
    $w=w_c$.  
    (b) The three-chain specific heat (${\cal C}_n$) per monomer with 
    the corresponding Boltzmann factor.  
    (c) Data collapse of energy.  (d) Data collapse of specific heat.  
}

   \label{fig:fig10}
 \end{figure}
}

\newcommand{\tripchainninew}{%
\begin{figure}[htbp]
   \centering
   \includegraphics[width=0.45\textwidth]{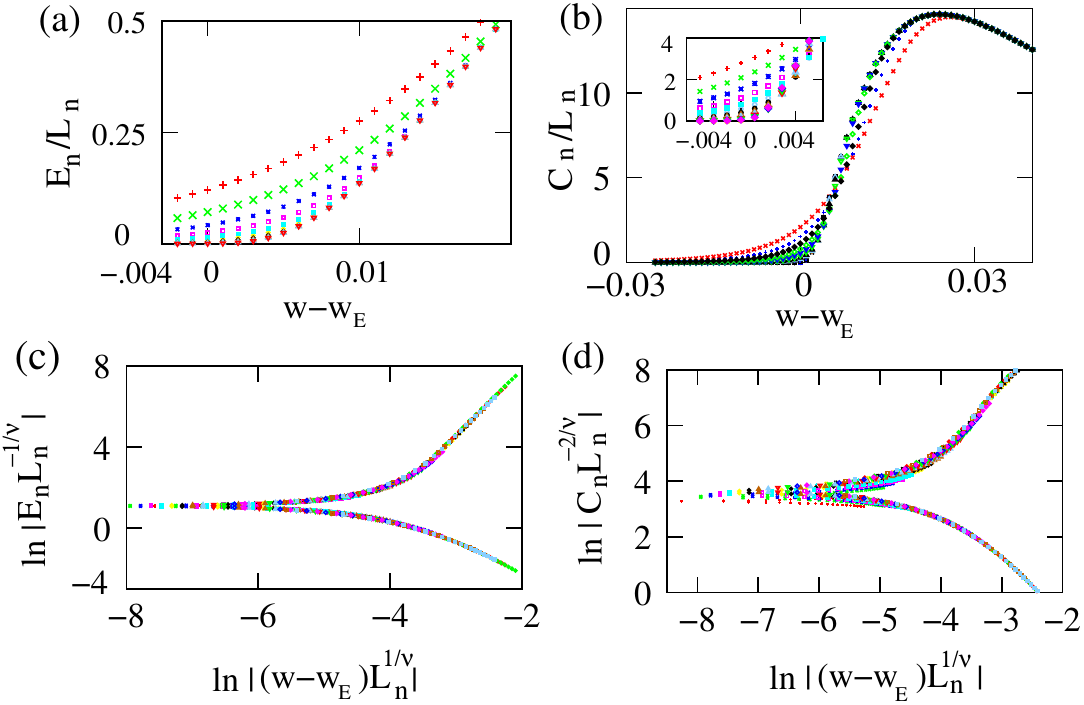}

   \caption{(Color online) For $b=9$.  (a) The three-chain average energy per monomer 
    versus the corresponding Boltzmann factor for chain length up to $2^{26}$ when 
    $y_c=b-1$.  The average energy shows a continuous transition at 
    $w=w_E$.  (b) The three-chain specific heat (${\cal C}_n$) per 
    monomer with the corresponding Boltzmann factor.  The length dependence is 
    shown in the inset. 
    (c) Data collapse of energy.  (d) Data collapse of specific heat. }

   \label{fig:fig11}
 \end{figure}
}

\newcommand{\energy}{%
\begin{figure}[htbp]
   \centering
   \includegraphics[width=0.45\textwidth]{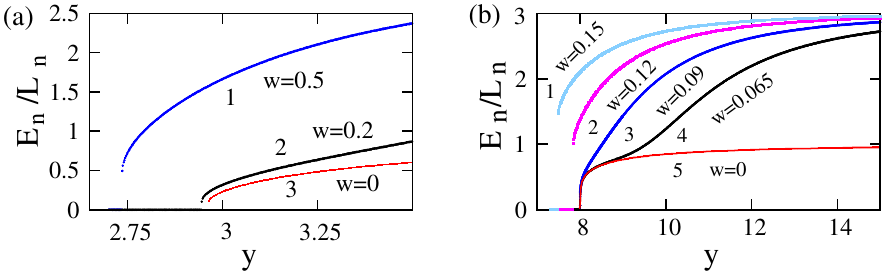}

   \caption{(Color online) The three-chain average energy per monomer with $y$ from 
    direct computation.  (a) For $b=4$, the average energy curves (marked 1, 2, and 
    3) with 
    the fixed values $w=0.5$, $w=0.2$, and $w=0$ show first order transitions.  
    (b) For $b=9$, the average energy curves (marked 1 and 2) with the fixed values 
    $w=0.15$ and $w=0.12$ show first-order transitions. Curves (marked 3, 4, and 5) 
    with the constant values $w=0.09$, $w=0.065$, and $w=0$ show a continuous 
    transition at $y_c=8$.  
}

   \label{fig:fig12}
 \end{figure}
}

\begin{document}
\title{Efimov effect of triple-stranded DNA: Real-space renormalization group and 
Zeros of the partition function}
\author{Jaya Maji}
\email{jayamaji@iopb.res.in}
\author{Somendra M. Bhattacharjee}
\email{somen@iopb.res.in}
\affiliation{Institute of Physics, Bhubaneswar-751005, India
}
\begin{abstract}
  We study the melting of three-stranded DNA by using the real-space
  renormalization group and exact recursion relations.  The prediction
  of an unusual Efimov-analog three-chain bound state, that appears at
  the critical melting of two-chain DNA, is corroborated by the
  zeros of the partition function.  The distribution of the zeros has 
  been studied in detail for various situations.  We show that the
  Efimov DNA can occur even if the three-chain ({\it{i. e.},} three-monomer) 
  interaction is repulsive in nature.  In higher dimensions,
  a striking result that emerged in this repulsive zone is a
  continuous transition from the critical state to the Efimov DNA.
\end{abstract}
\date{\today}

\maketitle

\section{INTRODUCTION}
In recent times the formation of triple-helical DNA has been a topic of 
considerable importance because of possible implications in the field of 
molecular biology.  In 1957, it was discovered that certain sequences of 
Watson-Crick double-helical DNA allow a third strand of DNA to bind via 
Hoogsteen or reverse Hoogsteen base pairing to form a triple 
helix \cite{rich,helen-1,helen-2}.  This triple helix formation has the 
potentiality to block transcription and thereby affect gene expression.  
Following this discovery, the experimental demonstration of the ability of 
a third chain to recognize the base sequences without the double-helical 
DNA revealing the base pairs renewed the interest in triple-helix DNA,  
especially its therapeutic applications \cite{jain,J-Clin}.  It is now known 
that not only DNA but even RNA \cite{rna} and PNA (polypeptide nucleic acid) 
are capable of forming a triple helix with duplex DNA \cite{pna-1,pna-2}.

Three-stranded DNA has been shown to exhibit an Efimov-like bound
state near or at the critical melting of duplex DNA \cite{efidna}.
The Efimov effect is the most striking phenomenon to occur in quantum
three-body systems with only two-body short-range pair
interactions \cite{efi-1,efi-2,efi-3,fonseca}.  An infinite number of 
bound states appear at the critical threshold of two body binding.  
There are several theoretical and experimental investigations using 
different models and methods that show this
effect \cite{efiexp-1,efiexp-2,efiexp-3,efirev}.  The universality of
this phenomenon encompasses the analogous classical model, namely, the
melting of three-stranded DNA \cite{efidna}.  An analogy is drawn
between the large quantum fluctuations near the zero-energy threshold
of two-body binding and the thermal Gaussian fluctuations at the
melting of duplex DNA.  
As discussed in Ref. \cite{efidna}, there is an exact mapping of the partition 
function of three ideal polymers with DNA base-pairing-type short-range 
interaction to the Green function of three-particle quantum mechanics 
under a transformation of the length of polymers to imaginary time.  
Furthermore, a scaling argument was used there to justify the occurrence of 
the effective two-chain attractive potential $\frac{1}{r^2}$ as a source of 
the Efimov effect.  Such a long-range interaction leads to a broad three-strand 
DNA bound state at or beyond the melting point of duplex DNA.  This is a state 
where no two strands are bound but the three are bound together.  We called this 
loosely bound state Efimov DNA \cite{bring}.  
This has also been observed from the renormalization group (RG) flows and exact 
numerical calculations for several model systems, in particular on hierarchical
lattices. 

Hierarchical lattices, by virtue of their discrete scaling, allow one to solve 
many models in statistical mechanics by exact renormalization group 
transformations \cite{smsmb,kaufgrif,haddad,berker}.  Furthermore, many 
approximate real-space RGs on real lattices can be viewed as exact real-space 
RGs on hierarchical lattices.  In the first study of the Efimov effect 
for Gaussian polymers, RG and exact numerics were used \cite{efidna}.  A part 
of our aim here is to analyze the Efimov phenomena exhibited by triple-stranded 
DNA from the classical phase transition point of view, especially by looking at 
the zeros of the partition functions.

Finding the zeros of a partition function in the complex plane of any physical 
variable is a mathematical way to understand and analyze phase transition 
phenomena.  However, finding those is often possible only for small sizes or 
soluble cases and not in general.  Yang and 
Lee first studied the Ising ferromagnetic system in a complex magnetic 
field to show that for a properly chosen variable the zeros lie on a unit 
circle, known as the Yang-Lee circle \cite{yanglee,leeyang}.  Later the zeros 
were studied in the complex temperature plane and other variables \cite{mfisher}.  
Since there cannot be any real zero, the zeros may accumulate and pinch the
real axis at a limit point in the thermodynamic limit.  This limit
point then identifies a transition point.  This method can provide
relevant information on phase transitions such as the critical field
or temperature and the values of the associated critical exponents.
Moreover, the distribution of zeros may form many complicated
structures other than a circle.  These structures are the separatrices
of the two types of flow to the two different stable fixed points of 
the RG transformation, and are similar to the Julia sets (see Appendix
\ref{sec:jset}) \cite{julia,derrida-1}.

In Ref. \cite{efidna} the RG flows were studied in the unbound region of 
the two- and the three-chain states.  By looking at the flows in the
unbound region of duplex DNA, where the chains are supposed to be
free, an effective three-chain bound state was predicted.  In this
paper we study the partition function of the three-chain system by
combining the recursion relations and the RG transformations, and then
finding the zeros.  We also extend the model to the three-chain
repulsive interaction regime.  In addition, we discuss several other
features of the zeros in the complex plane, for instance the detailed
structure, and the connection to the Julia set.

This paper is organized as follows.  In Sec.~\ref{model}, the three-polymer 
problem on a hierarchical lattice is introduced.  In Sec.~\ref{method}, the 
recursion relations from RG decimation and those for exact iterations are 
written.  The method of finding the zeros of the partition function is 
discussed, and we find the limit point of the zeros to locate the phase 
transition.  Section \ref{23chain} contains the results and discussions on 
the two- and the three-chain systems under different situations.  
In particular we estimate the transition point for Efimov DNA.  
Section \ref{nuevidence} extends the problem to three-chain repulsive 
interactions.  The existence of a transition between the Efimov DNA and 
the critical repulsive state in higher dimensions is established there.  
Appendixes \ref{sec:jset} and \ref{sec:limicycle} describe the Julia set and 
the limit cycle.

\section{MODEL}\label{model}
Let us consider the diamond hierarchical lattice as shown in
Fig.~\ref{fig:fig1}.  The lattice is generated iteratively by the
replacement of each bond at the $(n-1)^{\rm th}$ generation by a motif
of $\lambda b$ bonds to get the $n^{\rm th}$ generation, where
$\lambda$ and $b$ represent the bond scaling factor and the branching
factor, respectively.  The thermodynamic limit is obtained as $n\to\infty$ 
and in that limit the effective dimensionality of the lattice is 
\begin{equation}
d=\frac{\ln\lambda b}{\ln\lambda}.
\end{equation}
In this paper we shall choose $\lambda=2$.

\hlat
    \hlatint
One major feature about hierarchical lattices is their unusual scale
invariance property.  They have a discrete scaling symmetry.  That is
why an exact implementation of the real space RG technique is possible.  
The decimation of the $n^{\rm th}$ generation to arrive at the 
$(n-1)^{\rm th}$ generation is precisely what is needed in a RG 
transformation.  Once the partition function is known, it is possible to 
calculate the free energy and the other thermodynamic quantities.  One 
may even write down recursion relations for them. 

We consider three directed polymers on a diamond hierarchical lattice.
Three chains on the diamond hierarchical lattice are stretched from
bottom to top, but they can wander at intermediate points.  The
contact energies are defined at the bonds only.  The polymers are
assigned attractive potentials $-\epsilon$ and
$-\epsilon_{123}$ $(\epsilon,\epsilon_{123}>0)$ if a single bond is
shared by the two and the three polymers, respectively (see
Fig.~\ref{fig:fig2}).  At each generation, the length of each
polymer increases by a factor $\lambda=2$ so that the length of
polymers at the $n^{\rm th}$ generation is
\begin{equation}
  \label{eq:2}
  L_n=2^n.
\end{equation}

For the Efimov effect, just pairwise interaction is enough.  However in 
a RG procedure it is imperative to define the model with both 
$\epsilon$ and $\epsilon_{123}$, because the three-chain interaction
gets generated on a longer scale. 

\section{Method}\label{method}
\subsection{Renormalization group}\label{RSRG}
In this section we summarize the RG transformations and the exact
recursion relations for the partition functions.  The two ways of
handling the problem are just two different ways to look at it.  In
the RG case, we start from a large lattice and remove short scale
fluctuations by renormalizing the parameters, effectively reducing the
size of the lattice.  In contrast to this idea of thinning out the
degrees of freedom, in the second method the lattice is built
generation by generation so that one may study the behavior of any
quantity of interest as a function of the length of the polymers.  This is
useful in studying phase transitions because finite size scaling can
then be used to explore the nature of the transition.

We introduce the Boltzmann factors,
\begin{equation}
  \label{eq:1}
  y=\exp(\beta\epsilon),~\mathrm{and}~\ w=\exp(\beta\epsilon_{123}),
\end{equation}
where $\beta=1/k_BT$, $k_B$ being the Boltzmann constant and $T$ the
temperature.  The RG transformations of the two-chain and the three-chain 
Boltzmann factors are given by
 \begin{eqnarray}
 y{'}&=&\frac{(b-1)+ y^2}{b}\label{eq:a7},\label{eq:ry}\\
 w{'}&=&\frac{(b-1)(b-2)+ 3(b-1)y^2+ y^6w^2}{b^2{y{'}}^3},\label{eq:rw}
 \end{eqnarray}
where the primed variables $y{'}$ and $w{'}$ on the left hand side represent 
the renormalized values of the Boltzmann factors.  For details see 
Appendix {\ref{sec:recrela}}.  These recursion relations 
show that the three-body term is generated even though we start with 
$\epsilon_{123}=0$, {\it{i. e.}}, $w=1$.  As expected the three-chain 
interaction does not affect ({\it{i. e.}}, renormalize) the two-chain 
interaction.

For a given $y$ and $w$, the flows from successive use of
Eqs.~(\ref{eq:ry}) and (\ref{eq:rw}) would give us the phases and the
nature of the transitions.  One needs the fixed points for this
analysis.  For the two-chain system, the fixed points of $y$ are
(i) $y^*=1$, a stable infinite temperature fixed point representing an 
unbound state, and 
(ii) $y^*=(b-1)$, an unstable fixed point representing the two-chain melting 
or critical point.  In addition, (iii) $y^*=\infty$ (zero temperature,  
representing a bound duplex state) is an obvious stable fixed point, which 
does not come from the RG relation but from the RG flow.   
For a pure three-chain interaction ($y=1$) the fixed points of $w$ correspond to 
(i) $1$, infinite temperature,
(ii) $(b^2-1)$, an unstable, three-chain critical point, and 
(iii) $\infty$ (zero temperature), a stable fixed point, which comes from 
the RG flow.  
The two-chain melting is critical with a diverging length scale with
exponent \cite{smsmb}
\begin{equation}
\nu=\frac{\ln{\lambda}}{\ln\left(\left. \frac{dy{'}}{dy}\right|_{y\to{y_c}}\right)}\label{eq:nuy}
\end{equation}
and the specific heat exponent 
\begin{equation}
\alpha=2-\nu.\label{eq:alp}
\end{equation}
At the two-chain critical point $y_c=b-1$, the fixed points of $w$ are found 
to be
\begin{equation}
  w_{\pm}= \frac{b^2\pm\sqrt{4-24b+32b^2-12b^3+b^4}}{2(b-1)^3}.\label{eq:wpm}
\end{equation}
For $b=4$, $w_{\pm}=\frac{8}{27}\pm i\frac{\sqrt{23}}{27}$, are complex 
numbers.  In the range $2.303<b<8.596$ no real roots are found from the three 
chain RG relation [Eq.~(\ref{eq:rw})] for $y=y_c$.  These complex roots lead 
to a limit cycle behavior, which is intimately related to the Efimov 
effect (see Appendix \ref{sec:limicycle}).

\subsection{Exact recursion relations}\label{excal}
With the trace over all configurations the $n^{\rm th }$ generation
partition functions for single- ($C_n$), double- ($Z_n$), and triple-
($Q_n$) chain systems obey the recursion relations
\begin{eqnarray}
  C_{n}&=&bC^{2}_{n-1},\label{eq:b5}\label{eq:C}\\
  Z_n&=&b(b-1)C^4_{n-1}+bZ^2_{n-1},\label{eq:Z}\\
  Q_n&=&b(b-1)(b-2)C^6_{n-1}\nonumber\\
     &&
     +3b(b-1)C^2_{n-1}Z_{n-1}^2 +bQ^2_{n-1}.\label{eq:Q}  
\end{eqnarray}
The initial conditions are taken as 
\begin{equation}
  \label{eq:3}
C_0=1, \quad Z_0=y, \quad Q_0=y^3w.  
\end{equation}
The average energy and the specific heat are defined as 
\begin{equation}\label{eq:EC}
E_n=\frac{\partial\ln{\rm Q_n}}{\partial x},~~\mathrm{and}~~ 
{\cal C}_n=\frac{\partial{\rm E_n}}{\partial x},   
\end{equation}
where $x$ is the appropriate variable ($y$ or $w$ as the case may be).      
Although these definitions are different from the actual definitions, the 
proportionality factors are not crucial here.

For given $y$ and $w$, Eqs.~(\ref{eq:C})--(\ref{eq:Q}) give the partition 
functions for different $L_n$.  The average energy and the specific heat 
can be determined for different $L_n$ by writing down the recursion 
relations for derivatives of Eqs.~(\ref{eq:C})--(\ref{eq:Q}).   

\subsection{Zeros of the partition functions $Z_n$ and $Q_n$}\label{zerosofpf} 
If we take $w=1$, {\it{i. e.}}, no three-body interaction, then the
partition functions are polynomials in $y$.  In general, $Z_n$ is a
polynomial in $y$ of order $L_n$ while $Q_n$ is a multinomial in $y$
and $w$.  These partition functions are then completely described by
the zeros which are necessarily complex.  A phase transition is signaled by a real
limit point of the zeros.  However, the rapid growth of the order of
the polynomials makes it difficult to implement this program 
directly.  A different representation is used to get the
zeros \cite{derrida-1}.

By using the RG transformations of $y$ and $w$, the recursion relations from 
Eqs.~(\ref{eq:C})--(\ref{eq:Q}) can be reduced exactly to the forms 
\begin{eqnarray}
  Z_n(y)&=&b^{L_n}Z_{n-1}(y{'}),\label{eq:rZ}\\
  Q_n(y,w)&=&(b^{L_n})^{3/2}Q_{n-1}(y{'},w{'}),\label{eq:rQ}  
\end{eqnarray}
with $y{'}$ and $w{'}$ given by Eqs.~(\ref{eq:ry}) and (\ref{eq:rw}).    
These relations can be verified by direct substitution and, if necessary, by 
the method of induction.

Since the zeros determine a polynomial completely, the two-chain partition 
functions can be written as 
\begin{eqnarray}
Z_n(y)&=&b^{L_n-1}{\displaystyle\prod_{l=1}^{L_n}}{(y-q_l)},\label{eq:Zpoly1}\\
\mathrm{and}~~~Z_{n-1}(y)&=&b^{L_{n-1}-1}{\displaystyle\prod_{j=1}^{L_{n-1}}}{(y-\tilde q_j)},\label{eq:Zpoly2}
\end{eqnarray}
where the $q_l$'s and $\tilde q_j$'s are the zeros of the partition functions 
$Z_n(y)$ and $Z_{n-1}(y)$, respectively.  These zeros appear in complex-conjugate 
pairs.  With the substitution of Eqs.~(\ref{eq:Zpoly1}) and (\ref{eq:Zpoly2}), 
Eq.~(\ref{eq:rZ}) becomes
\begin{equation}
b^{L_n-1}{\displaystyle\prod_{l=1}^{L_n}}{(y-q_l)}=b^{L_n}b^{L_{n-1}-1}
{\displaystyle\prod_{j=1}^{L_{n-1}}}(y{'}-\tilde q_j).\label{eq:Zsub}
\end{equation} 
Then the use of Eq.~(\ref{eq:ry}), the relation between $y'$ and $y$, gives two 
roots from each factor on the right hand side, so that the $q_l$'s are the 
solutions of
\begin{equation}
\frac{(b-1)+ y^2}{b}=\tilde q_j,\label{eq:q}
\end{equation} 
{\it{i. e.}},
\begin{equation}
q=\pm\sqrt{b\tilde q_j-(b-1)}.\label{eq:qroot}
\end{equation} 
The subscript of $q$ is omitted.  
This clearly shows that if we know the $2^{n-1}$ zeros $\tilde q_j$ of $Z_{n-1}(y)$, we 
will be able to know the $2^n$ zeros $q_l$ of $Z_n(y)$.   One may start with the roots 
of $Z_1$ and generate successively the roots of each generation, by just solving a 
quadratic equation.   
  
Instead of generating all the roots, a random generation is more easily implementable.  
With an initial value $y_0$ chosen randomly from the two roots of $Z_1$, the new  
roots are determined by Eq.~(\ref{eq:qroot}).  If one of them is chosen at random and 
substituted as $\tilde q_j$, the roots for the  next generation can be found.  Thus, 
after the $n$th iteration, the set obtained is basically the zeros in the complex 
$y$-plane.  These roots are nothing but the zeros of the partition function found from 
different sizes of the lattice, which in this problem would be equivalent to  different 
lengths of polymers.  The zeros quickly converge and as $n\to\infty$ we look for the 
limit point on the real axis.  Apart from that, the distribution in the complex 
$y$-plane itself is of interest.
This method has been generalized for the three-chain system.  

\section{Behavior of zeros: two-chain and three-chain systems}\label{23chain}
\subsection{Two-chain system: $b=4$}\label{2chain}
For different branching factors, fractal-like structures are obtained from the zeros 
of the partition functions of the two- and the three-chain systems.  
We considered only $b=4$ as a representative of the range where there is no real 
fixed point along the two-chain critical line.
        \bfderridup
For $b=4$ the structure shown in Fig.~\ref{fig:fig3}(a) is obtained in the complex 
$y$ plane from the exact recursion relation Eq.~(\ref{eq:rZ}).  
Exact solutions are possible only 
up to the $n=6$ generation because of computational hardware limitations.  
This is insufficient, as the thermodynamic limit ($n\to\infty$) is needed to observe a 
phase transition.  
Finding zeros at random from the RG relations [Eqs.~(\ref{eq:ry}) and (\ref{eq:rw})] 
overcomes such difficulties and hence large lengths can be reached.    
The zeros obtained from Eq.~(\ref{eq:q}) give the fractal-like structure shown in 
Fig.~\ref{fig:fig3}(b).  The  accessed zero nearest to the real axis  
approaches the two-chain transition point $y_c=3$ for large $n$.  
     \figbfourfrac
Apart from the limit point, the distribution of the zeros in the
complex $y$ plane is also non-trivial.  

The first feature to note is that the zeros do not seem to lie on a
smooth differentiable curve.  A zoomed picture of a small
cross section of the  structure for the two-chain system [from
Fig.~\ref{fig:fig3}(b)] is shown in Fig.~\ref{fig:fig4}(a).
Further the selected regions have been zoomed successively and are shown in
Figs.~\ref{fig:fig4}(b) and \ref{fig:fig4}(c). The self-similarity of the
structure is visible.  This is an indication of  the fractal nature of
the distribution.  Further analysis required for a quantitative description 
is not done here.

These fractal like structures obtained above are nothing but the separatrices 
of the set of RG flows in the complex plane to the appropriate stable fixed 
points.  These separatrices for iterations of any function in the 
complex plane are known as the Julia set (see appendix \ref{sec:jset}).  
The sets are obtained after an infinite number of iterations of a recursive formula 
by identifying the points that do not flow to the stable fixed points.  Our method 
of finding the zeros by using the RG relations is in fact equivalent to an inverse 
iteration method, which is more efficient in producing such structures.  

In Fig.~\ref{fig:fig5}(a) the RG flows are shown in the complex $y$ plane
for a two-chain  system.  The dotted line (red curve) shows the
flow towards the stable fixed point $y=1$, {\it{i. e.}}, the high
temperature region, when we start with a value from the inner region
of the fractal-like structure.  On the other hand, a point from the
outskirts of the line of zeros flows to the stable fixed point
$y=\infty$, which is the bound state with zero temperature.  The
critical point, being an unstable fixed point, does not actually belong
to the set but, as discussed, is a limit point --- in a sense a
boundary of the set.

The second feature to note is the 3-like shape near the real-axis limit point.  
It is not arbitrary.  
The angle at the limit point in the complex plane is related to the specific heat 
exponent by \cite{itzykson} 
\begin{equation}\label{eq:phi}
\tan{(\phi\nu)}=-\tan{(\pi\alpha)}+\frac{A_-}{A_+}{\rm csc}{(\pi\alpha)},
\end{equation}
where $\phi$ is the angle between the tangent of zeros at the limit point, and the 
real axis of $y$, and $A_{\pm}$ are the amplitudes of the specific heat on the low 
and the high $y$ side of the transition. Just like the exponents, $A_{-}/A_{+}$ 
is a universal number for a universality class of transition.  For the two-chain 
problem, we know that $A_{-}/A_{+}\to\infty$ as $A_{+}=0$.  Therefore the angle 
$\phi$ is given by   
\begin{equation}\label{eq:phiang}
\phi=\frac{\pi}{2\nu}.
\end{equation} 
     \bfdupflow
The zeros obtained by the successive iterations of the one close to the real  
axis are shown in Fig.~\ref{fig:fig5}(b) by the triangles.  They approach the real 
axis in a linear fashion with an angle $\phi$, given by the straight line  
\begin{equation}\label{eq:line}
{\rm Im}~z=({\rm Re}~z-c)\tan{\frac{\pi}{2\nu}},
\end{equation}
in the generic complex $z$ plane with $\nu$ from Eq.~(\ref{eq:nuy}).  Here $c$ 
represents the limit point of the zeros on the real axis.  
The zeros occur in complex conjugate pairs.  Therefore if we take the mirror image 
of the distribution of zeros about the real axis in Fig.~\ref{fig:fig5}(b), the 
beak of the 3-like shape can be obtained.   

   \bftripflow
\subsection{Three chain system: $b=4$}\label{3chain}
We have calculated the zeros of $Q_n(1,w)$ for a three-chain system with a pure 
three-chain interaction.  By considering $y=1$ in Eq.~(\ref{eq:rw}), we get 
\begin{equation}
w{'}=\frac{(b^2-1)+w^2}{b^2}.
\end{equation}
The zeros come from the equation $$q_l=\pm\sqrt{b^2\tilde q_j-(b^2-1)},$$ 
where the $q_l$'s and $\tilde q_j$'s are the zeros of $Q_{n}(1,w)$ and $Q_{n-1}(1,w)$, 
respectively.  
The distribution of zeros is the Julia set which has a fractal-like structure shown in 
Figs.~\ref{fig:fig7}(a),\ref{fig:fig7}(c), and \ref{fig:fig7}(d).  By choosing the 
zero near to the limit point $w_c$, the nature of the distribution can be determined, 
as shown in Fig.~\ref{fig:fig7}(b) by the straight line given by Eq.~(\ref{eq:line}) 
with $\nu$ of Eq.~(\ref{eq:nuw}) and $c=w_c$.          

\subsection{Efimov DNA: $b=4$}\label{3chainefimov}
The idea is to show the Efimov transition point of DNA by finding the limit point 
of zeros on the real $y$ axis.  Although we consider $w=1$, the effective 
three-chain interaction develops by renormalization.  As a result the zeros 
found from Eqs.~(\ref{eq:C})-(\ref{eq:Q}) seem to pinch the Re($y$) axis at a point 
where no pair of chains is bound.  The exact solutions are shown in 
Fig.~\ref{fig:fig6}(a) for $n=6$.  On a finer scale the zeros are 
shown in Fig.~\ref{fig:fig6}(b).  For such small lattices  the limit point is not 
accessible, hence an extrapolation scheme may be used.  The zeros nearest to the 
Re$(y)$ axis, obtained in different generations ($n=2,...,6)$ are 
shown in Fig.~\ref{fig:fig6}(c) by black dots.  A straight line nicely fits 
these zeros and is shown by the solid red curve.    

    \figbfour
The straight line intersects the real axis at $y=2.321$. This value is the large $n$ 
extrapolation and can be taken as an estimate of the Efimov transition.  We may 
compare this extrapolated value with the previous RG-based estimate of $y_E=2.32402$.  
Finding the zeros for the two-chain system is easier than for the three-chain system.  
Since the three-chain equation holds both the variables $y$ and $w$, finding 
zeros from the three-chain RG relation is tantamount to generating the full relation 
for $Q_n$.  This is because one needs to keep $w$ at all the intermediate values of 
$n$ and then, at the the desired value of $n$, $w$ is to be set to $1$.  One 
sees the difficulty of the Efimov physics eventhough $w=1$. 
It is tempting to simplify the recursion relation at the cost of some approximation.  
We set $w=w{'}=1$ to get a renormalized $y{'}$ that describes the three-chain 
system.  Such a relation follows from Eq.~({\ref{eq:rw}}), as
\begin{equation}
{y{'}}^3=\frac{(b-1)(b-2)+ 3(b-1)y^2+ y^6}{b^2}.\label{eq:w1}
\end{equation}
The zeros obtained from Eq.~(\ref{eq:w1}) spread out in a ``Milky Way'' over a 
region in the complex plane of $y$.  The spread makes it difficult to make an 
estimate of the real-axis limit point, but one may use the width to put a bound 
on the Efimov transition point [see Fig.~\ref{fig:fig6}(c)].        

\subsection{Efimov DNA at ${\bf y_c=b-1}$: $b=9$}\label{critline}
A study along the critical threshold of the two-chain melting is
quite interesting.  No real fixed point for $w$ exists for
Eq.~(\ref{eq:rw}) when $b$ is in the range $2.303\le b\le 8.596$ along the
$y=y_c$ line.  For $y=y_c$, the single parameter RG relation is 
\begin{equation}
w{'}=\frac{(b-2)+ 3(b-1)^2+ (b-1)^5w^2}{b^2(b-1)^2}.
\end{equation}
The two fixed points for this case are given by Eq.~(\ref{eq:wpm}).  
For $b=9$, these are
\begin{subequations}
\begin{align}
w=w_s=0.0655347...~~~ &{\rm (stable),}\label{eq:ws}\\
w=w_E=0.0926684...~~~ &{\rm (unstable).}\label{eq:wE}  
\end{align}
\end{subequations}
The unstable fixed point, as the phase transition point, determines the
limit point of the zeros of the partition function on the real axis.
Hence it can be predicted that at the two-chain melting point, by tuning 
$w$, a transition occurs at $w=w_E$, from the Efimov DNA to the critical 
state of polymer pairs.  Figure \ref{fig:fig8}(a) shows the distribution of zeros 
of $Q_n(y_c,w)$ in the complex $w$ plane.  The set of these zeros is a Julia set, 
separating the flows to the stable fixed points.  The stable fixed point in the 
inner region of the set is given by Eq.~(\ref{eq:ws}).  
The zeros near the real axis approach $w=w_E$ linearly, following Eq.~(\ref{eq:line}) 
with $c=w_E$ and $\nu$ of Eq.~(\ref{eq:nuyc}) as shown in  Fig.~\ref{fig:fig8}(b).  
A detailed discussion is given in the next section.
     \figbnine

\section{Efimov DNA: RG flow and numerical evidence}\label{nuevidence}
To explore the robustness of the Efimov effect, we now include a three-chain 
repulsive interaction along with the pairwise attractive one.  
The three-chain interaction is attractive when $w>1$ and repulsive 
for $0\le w<1$.  
For $w=0$, representing the hard core three-chain repulsive interaction, three 
chains can never be on the same bond in this model.

\subsubsection{$b=4$}\label{wlt1}
For $b=4$ the RG phase diagram is shown in Fig.~\ref{fig:fig9}(a).  The solid red 
line is the separatrix connecting the pure three-chain transition point $(1,w_c)$ 
to an Efimov transition point for $w=0$.  Each point on the solid line represents 
an Efimov transition point.  In other words keeping $w$ fixed, by changing $y$, we 
can see a melting of a loosely bound Efimov DNA with no pairwise binding.
      \phasedia
The region enclosed between this separatrix (solid red line) and the $y_c=3$  
line is the Efimov region and $(y,w)$ flows to $(1,\infty)$.  Below the 
solid red line is the high temperature zone of denatured DNA, where RG flows 
are to $(1,1)$.  The region to the right of the $y_c=3$ line is the two-chain 
bound state.  The area below the dashed curve, where the RG flow takes $w$ to zero 
when two-chain pairs are strongly bound, represents a different state 
where one finds a three-chain bound state but with no three-chain contact.  
The dashed line is then a crossover line.  It remains to be seen if under 
some conditions this crossover line becomes a true phase transition line.     
      \energy

\subsubsection{$b=9$, $y_c=b-1$}\label{wlt12}
The RG phase diagram is shown in Fig.~\ref{fig:fig9}(b) for $b=9$.  In the 
diagram two separatrices (the solid red line and the dashed line) 
meet at an unstable fixed point.  The two fixed points $w=w_s$ and $w=w_E$ are 
shown in Fig.~\ref{fig:fig9}(b).  The presence of any unstable fixed point reflects 
a continuous transition along the two-chain critical line.  Hence we can say that 
by tuning the three-chain repulsive interaction parameter or temperature in the 
repulsive zone a transition can be induced in the Efimov DNA at the critical 
threshold of duplex  binding.  The transition is from the Efimov state to the 
critical state of pairs dominated by the three-chain repulsion.  The Efimov region 
is now restricted by a separatrix connecting the two unstable fixed points 
$(1,w_c)$ and $(y_c,w_E)$ and the critical line $y_c=b-1$.  

On the critical line at both the fixed points $w=w_s$ and $w=w_E$, $y$ is 
a relevant variable (unstable in the $y$ direction).  But $y$ does not couple to $w$ 
in the RG equation [Eq.~(\ref{eq:ry})].  The melting for $w<w_E$ would be similar 
to the pure two-chain melting described by Eqs.~(\ref{eq:nuy}) and (\ref{eq:alp}).  
In the $y$-$w$ plane, ($y_c$,$w_E$) is a multicritical point where the line of 
first-order transitions goes over to a line of critical points.

\subsubsection{Data collapse}
We now provide  numerical evidence for the above RG-based inferences.  Exact 
numerical calculations of the average energy and the specific heat are done by 
iterating the partition functions and their higher derivatives for lattices 
of various sizes for  different fixed values of $w$.   
Figure \ref{fig:fig12}(a) for $b=4$ shows that at $w=0.5$, $w=0.2$, and $w=0$,   
there are first-order transitions.  The transition points estimated from 
the point of discontinuity are shown by the filled circles 
in Fig.~\ref{fig:fig9}(a).  They are on the separatrix and are the Efimov 
transition points for the corresponding values of $w$.  

The energy curves in Fig.~\ref{fig:fig12}(b) for $b=9$ with  $w=0.15$ and $w=0.12$, 
show first-order transitions.  These transition points are shown by the filled 
circles in Fig.~\ref{fig:fig9}(b).  In contrast, the energy curves (marked 3, 4, and 
5) show continuous transitions for $w=0.09$, $w=0.065$, and $w=0$, respectively at 
$y_c=8$.  This is consistent with the RG prediction of Fig.~\ref{fig:fig9}(b).

The energy and the specific heat curves are shown in 
Figs.~\ref{fig:fig10}(a) and \ref{fig:fig10}(b) for $b=9$, $y=1$ and in 
Figs.~\ref{fig:fig11}(a) and \ref{fig:fig11}(b) for $b=9$, $y_c=b-1$.        
Also the corresponding finite size scaling is shown in 
Figs.~\ref{fig:fig10}(c) and \ref{fig:fig10}(d) for $b=9$, $y=1$ and in 
Figs.~\ref{fig:fig11}(c) and \ref{fig:fig11}(d) for $b=9$, $y_c=b-1$.  
The finite size scaling behavior of different thermodynamic quantities is described 
by the length scale exponents.  In analogy with Eq.~({\ref{eq:nuy}}), 
the exponents to describe the three-chain transition for $y=1$ and $y_c=b-1$ at 
appropriate critical points are given by  
      \tripchainnine
\begin{eqnarray}
\nu&=&\frac{\ln{2}}{\ln{\frac{2(b^2-1)}{b^2}}},\label{eq:nuw}\\
\nu&=&\frac{\ln{2}}{\ln\left(\left. \frac{\partial w{'}}{\partial w}
\right|_{\substack{y_c={b-1}\\
w\to w_E}}\right)}.\label{eq:nuyc}
\end{eqnarray}
Around a critical point  one should see a  finite size scaling.
Therefore the average energy and the specific heat obeying the finite
size scaling can be written in the forms
\begin{eqnarray}
E&\sim&L^{1/\nu}f(L^{1/\nu}|w-w^{*}|),\label{eq:sfE}\\
\cal C&\sim&L^{2/\nu}f(L^{1/\nu}|w-w^{*}|),\label{eq:sfC}
\end{eqnarray}
with appropriate $\nu$ and $w^*$.  
      \tripchainninew
In Figs.~\ref{fig:fig10}(c) and \ref{fig:fig10}(d) we see that the average energy 
and the 
specific heat scale as $E_nL_n^{-1/\nu}$ and ${\cal C}_nL_n^{-2/\nu}$, 
respectively, when plotted versus $|(w-w^{*})|L_n^{1/\nu}$ with the $\nu$ of 
Eq.~({\ref{eq:nuw}}) and $w^{*}=w_c$ for $y=1$, all the data collapse onto a single 
curve for different lengths of polymers, where $n=6,7,...,26$.

Figures \ref{fig:fig11}(c) and \ref{fig:fig11}(d) show similar plots for the critical 
line ($y_c=b-1$) with $\nu$ of Eq.~(\ref{eq:nuyc}) and $w^{*}=w_E$.   
Since the specific heat diverges with increasing length, data collapse is good for 
the case $y=1$.  The data collapse for the case $y_c=b-1$ is not so good due to 
a smoother  behavior of the specific heat at the critical point.  
These establish the weak criticality at $w=w_E$.  

\section{Summary}
To summarize, the RG relations and exact recursion relations are used to study 
the three-chain system on a diamond hierarchical lattice.  Our emphasis 
is on  the Efimov-like state exhibited by the three-chain system at or beyond the 
two-chain melting, where no two chains are bound, and the nature of the transitions.  
Fractal-like structures are obtained for the zeros of the partition functions.  
These zeros, when they pinch the real axis, determine the phase transition points.  
We find that all the transition points obtained from RG flows, are in good agreement 
with the zeros of the partition function on the real axis.  The Efimov transition point 
thus found strengthens the prediction of Efimov-like phenomena for the three-chain 
system.  We have shown that the Efimov effect is exhibited by a three-chain 
system even if there is a repulsive three-chain interaction.  A transition can 
be induced in higher dimensions from the Efimov state to the three-chain critical 
repulsive state at the melting of duplex DNA.  The transition to this three-chain 
critical repulsive state is continuous and obeys a finite size scaling law with 
exponents obtained from the RG.  In the $(y,w)$ phase diagram, $(y_c,w_E)$ is a 
multicritical point.  

Although the model studied in this paper is simplistic, mainly to get exact results, 
still the denaturation transition induced by bubble formation accompanied by 
diverging length scales is the generic scenario for more realistic polymeric 
models.  The qualitative picture is therefore expected to be valid for those models 
too.  We await experimental evidence for the existence of the Efimov DNA or the 
Efimov transition.  Again, the existence of such a state remains a challenge for 
molecular dynamics and Monte Carlo simulations.
  
\begin{acknowledgements}
J.M. would like to thank Professor A. Khare for discussions on Julia sets.
\end{acknowledgements}

\appendix
\section{RG relations}\label{sec:recrela}
The configurations of the two-chain system on a motif of a hierarchical 
lattice can be classified as two independent chains or inherently two-chain 
configurations as shown in Fig.~\ref{fig:fig2}(c).  By summing over all 
configurations the partition functions 
for $n=0$ and $n=1$ generation lattices for general $b$ can be written 
as \cite{smsmb,efidna} 
\begin{eqnarray}
Z_0(y)&=&y,\\
Z_1(y)&=&b(b-1)+by^2.
\end{eqnarray}
In RG decimation, $2b$ bonds of the $n=1$ generation are replaced by a single 
bond at the $n=0$ generation.  Then RG demands
\begin{equation}\label{propcon}
Z_0(y{'})\propto Z_1(y),
\end{equation}
where $y{'}$ is the renormalized Boltzmann factor.  With the free chain boundary 
conditions ({\it i. e.}, $y=1$, implies $y{'}=1$), the proportionality constant of 
Eq.~(\ref{propcon}) can be determined.  The RG transformation for the two-chain 
Boltzmann factor then becomes  
\begin{equation}
y{'}=\frac{b(b-1)+ by^2}{b^2}.
\end{equation}
The RG relation for the three-chain case can also be written in the same spirit 
as in the two-chain case.  The free chain condition is that $y=w=1$ implies 
$y'=w'=1$.  It is also to be noted that when three chains share the same bond the 
contribution is $y^3w$.  The RG transformation for $w$ is then 
\begin{equation}
y{'}^3w{'}=\frac{b(b-1)(b-2)+3b(b-1)y^2+by^6w^2}{b^3}, 
\end{equation}
where $w{'}$ is the renormalized value of $w$.

\section{Julia set}\label{sec:jset}
The standard definition of a Julia set is the set of points on the complex plane 
which flow to a fixed point (no divergence) after a function, {\it{e. g.},}  
\begin{eqnarray}
z_{n}&=&z_{n-1}^2+c,\label{Eq:f1}
\end{eqnarray}
is repeatedly applied, where $c$ is any arbitrary constant, and could be real or 
complex.  Let us choose $c=0$.  
The fixed point solutions for $c=0$ are $z=0, 1, \infty$, where $z=1$ is the
unstable fixed point.  Here, for $n\to\infty$, $z_{n+1}\to 0$, when we start 
with $|z_0|<1$ and $z_{n+1}\to\infty$, when we start with $|z_0|>1$.
Therefore the unit circle $|z|=1$ is the boundary between the two stable
fixed points $z=0, \infty$.  The unstable point lies on this boundary.

\section{Limit Cycle}\label{sec:limicycle}
For two successive generations Eqs.~(\ref{eq:rw}) will be  
\begin{equation}
w_{n}-w_{n+1}=f(w_{n+1})-w_{n+1}\label{Eq:difeq}.
\end{equation}
But if the continuum limit is taken, Eq.~(\ref{Eq:difeq}) can be written as 
\begin{equation}
l\frac{dw}{dl}=-(w-w_+)(w-w_-),\label{Eq:dif}
\end{equation}
at the critical line $y_c=b-1$, where $l=\ln L$ and $L=2^n$.  For complex 
$w_{\pm}=\alpha\pm i \beta$, the solution of Eq.~(\ref{Eq:dif}) is then  
\begin{equation}
w=\alpha-\beta\tan\beta(\ln l+\theta)\label{Eq:wlim},
\end{equation}
where $\theta$ is the integration constant.
The above equation reflects the periodicity of $w$ in $\ln l$ with the property
\begin{equation}
w(l)=w(l\lambda),\ {\rm where}\  \ln\lambda=\frac{\pi}{\beta}.\label{Eq:cond}
\end{equation}
Here as $l$ increases $w$ approaches $\pm\infty$.  This behavior can be mapped into 
a limit cycle in the complex plane with a phase factor defined by the equation  
\begin{equation}
e^{i\phi}=\frac{w-w_+}{w-w_-}.\label{phase}
\end{equation}
With the help of Eq.~(\ref{Eq:dif}) and its derivative, $\phi$ will be  
\begin{equation}
\phi=\frac{\beta}{\alpha}\ln l+\phi_0, 
\end{equation}
where $\phi_0$ is the integration constant.  

Our model on the hierarchical lattice is a discrete model.  Certainly
a limit cycle is  obtainable from the RG relations in the continuum
limit, but it is not straight forward to do so in the discrete case.



\begin{thebibliography}{99}
\bibitem{bring} Efimov DNA is not like Borromean rings.  Borromean rings remain 
         linked even in the denatured state of DNA because of their closed form.  
         On the other hand the closed form of DNA is not a necessity for Efimov DNA.
\end{thebibliography}
\end{document}